\documentclass[pra,twocolumn,showpacs,superscriptaddress,amsfonts,amsmath,floatfix,nofootinbib]{revtex4}

\usepackage{graphicx}

\newcommand{\Journal}[4]{#1 {\bf #2}, #3 (#4)}
\newcommand{\PR}{Phys. Rev. }
\newcommand{\PRL}{Phys. Rev. Lett. }

\newcommand{\PRB}{Phys. Rev. B }
\newcommand{\JMP}{J. Math. Phys. }

\newcommand{\PLA}{Phys. Lett. A }

\newcommand{\Exp}[1]{\mathrm{e}^{#1}}

\begin{document}

\title{Generalized harmonic-fluid approach for the off-diagonal correlations of a one-dimensional interacting Bose gas} 
\author{N. Didier} 
\email{Nicolas.Didier@grenoble.cnrs.fr} 
\affiliation{Universit\'e Joseph Fourier, Laboratoire de Physique et de Mod\'elisation des Milieux Condens\'es, C.N.R.S. B.P. 166, 38042 Grenoble, France} 
\author{A. Minguzzi} 
\email{Anna.Minguzzi@grenoble.cnrs.fr} 
\affiliation{Universit\'e Joseph Fourier, Laboratoire de Physique et de Mod\'elisation des Milieux Condens\'es, C.N.R.S. B.P. 166, 38042 Grenoble, France} 
\author{F.W.J. Hekking} 
\email{Frank.Hekking@grenoble.cnrs.fr} 
\affiliation{Universit\'e Joseph Fourier, Laboratoire de Physique et de Mod\'elisation des Milieux Condens\'es, C.N.R.S. B.P. 166, 38042 Grenoble, France}

\begin{abstract}
We develop a generalized harmonic-fluid approach, based on a regularization of the effective low-energy Luttinger-liquid Hamiltonian, for a one-dimensional Bose gas with repulsive contact interactions. 
The method enables us to compute the complete series corresponding to the large-distance, off-diagonal behavior of the one-body density matrix for any value of the Luttinger parameter~$K$. 
We compare our results with the exact ones known in the Tonks-Girardeau limit of infinitely large interactions (corresponding to~$K=1$) and, different from the usual harmonic-fluid approach, we recover the full structure of the series. 
The structure is conserved for arbitrary values of the interaction strength, with power laws fixed by the universal parameter~$K$ and a sequence of subleading corrections. 
\end{abstract}
\maketitle

\section{Introduction}

Quasi-one-dimensional quantum fluids display unique properties due to the major role played by quantum fluctuations in reduced dimensionality (see \textit{e.g.}~\cite{Giamarchi_book}). 
For example, in an interacting 1D Bose fluid quantum fluctuations destroy the off-diagonal long-range order of the one-body density matrix (or first-order coherence function), defined as $\rho_1(x_1,x_2)=\langle \psi^\dagger(x_1)\psi(x_2)\rangle$, where $\psi(x)$ is the bosonic field annihilation operator at position~$x$. 
In contrast to its 3D counterpart, where the one-body density matrix at large distances tends to a constant which corresponds to the fraction of Bose-condensed atoms~\cite{PenroseOnsager}, in 1D the one-body density matrix at zero temperature decays as a power law~\cite{Luther,Haldane81}, the coefficient of the power law being fixed by the interaction strength: the decay is faster as the interaction strength increases from the quasi-condensate regime with a fluctuating phase~\cite{Popov_book}, to the Tonks-Girardeau regime~\cite{Girardeau60} of impenetrable bosons. 

The one-body density matrix is not only a fundamental quantity as it measures the 
macroscopic coherence properties of a quantum fluid with bosonic statistics, but also 
because it is directly related (by Fourier transformation) to the momentum distribution 
of the fluid. While the momentum distribution for a Bose fluid is typically narrow and 
peaked around wavevector~$k=0$, its specific form, its broadening due to interactions 
and its possible singularities give a wealth of information about the nature of the 
correlated fluid under study. 
 
Quasi-one-dimensional Bose fluids find an experimental realization in experiments with 
ultra-cold atomic gases confined to the minima of a 2D optical lattice~\cite{1D_exps}. 
For those systems, the momentum distribution is one of the most common observables and 
the one-body density matrix has been measured as well~\cite{esslinger}, though not yet in 
the quasi-1D geometry. 
 
From a theoretical point of view, the calculation of the one-body density matrix for the 
1D interacting Bose gas has a long history. In the Tonks-Girardeau limit where the exact 
many-body wavefunction is known by means of a mapping onto a gas of spinless fermions~\cite{Girardeau60}, 
the problem reduces to the evaluation of a $(N-1)$-dimensional integral. This 
mathematical challenge has been addressed first by Lenard~\cite{Lenard}, and by several 
subsequent works (see \textit{e.g.}~\cite{Vaidya_Tracy,Jimbo,Bogoliubov,Gangardt04}). The main 
result is the evaluation of the large-distance behavior of the one-body density matrix in 
the form of a series expansion, (from~\cite{Gangardt04}) 
\begin{eqnarray} 
\label{eq:rhoTG} 
\rho_1^{\mathrm{TG}}(z)&=&\frac{\rho_\infty}{|z|^{1/2}} 
\left[1-\frac{1}{32}\frac{1}{z^2} -\frac{1}{8}\frac{\cos(2 
  z)}{z^2}- \frac{3}{16}\frac{\sin (2 z)}{z^3}\right.\nonumber \\ &&\left.+ 
\frac{33}{2048}\frac{1}{z^4}+\frac{93}{256}\frac{\cos(2z)}{z^4}+....\right], 
\end{eqnarray} 
where the constant~$\rho_\infty$ and the coefficients have been calculated exactly. In 
Eq.~(\ref{eq:rhoTG}) and in the following we express the one-body density matrix in units 
of the average particle density~$\rho_0$ and as a function of the scaled relative 
coordinate~$z=\pi \rho_0(x_1-x_2)$. 
 
For the case of arbitrary interaction strength, the calculation of the correlation 
functions remains a challenge (see \textit{e.g.}~\cite{Korepin_book}), although the model of bosons 
with contact repulsive interactions is integrable by the Bethe-Ansatz 
technique~\cite{LiebLiniger}. The power-law decay at large distances can be inferred using a 
harmonic-fluid approach~\cite{Tomonaga50,MattisLieb65,EfetovLarkin75,Luther,Haldane81}. 
The latter is based on an effective, low-energy Hamiltonian describing the 
long-wavelength collective excitations of the fluid having a linear excitation spectrum 
$\omega(k)=v_s k$. The resulting structure for the large-distance series of the one-body 
density matrix reads (from~\cite{Haldane81}) 
\begin{equation} 
\label{eq:rhoLL} 
\rho_1^{\mathrm{LL}}(z)\sim \frac{1}{| 
 z|^{1/2K}}\sum_{m=0}^\infty B_m \frac{\cos(2 m z)}{z^{2m^2K}}, 
\end{equation} 
where $K=\pi \rho_0/mv_s$ is the universal Luttinger-liquid parameter and 
 the coefficients~$B_m$ are nonuniversal and cannot be obtained by the 
 harmonic-fluid approach. 
By noticing that the harmonic-fluid approach is valid also in the Tonks-Girardeau regime 
and corresponds to the case of Luttinger parameter~$K=1$, one can directly compare the 
predictions of the two methods.
Specifically, this comparison shows that the structure of Eq.~(\ref{eq:rhoTG}) is richer than that of Eq.~(\ref{eq:rhoLL}) obtained by the standard harmonic-fluid approach.

The central result of this work is that by defining a properly regularized 
 harmonic-fluid model (to be detailed below) the following general structure of 
 the one-body density matrix can be obtained {\em for arbitrary values of the Luttinger 
 parameter}~$K$ : 
\begin{eqnarray} 
\label{eq:rhoGLL} 
\rho_1 (z)\!\!\!&\sim& \!\!\!\frac{1}{| 
 z|^{1/2K}}\left[1+ \!\!\!\sum_{n=1}^\infty \frac{a'_n}{z^{2 n}}+ 
 \!\!\!\sum_{m=1}^\infty b_m \frac{\cos(2 m z)}{z^{2m^2K}} 
 \left(\sum_{n=0}^\infty \frac{b'_n}{z^{2 n}}\right)\right. \nonumber 
 \\\!\!\! && \!\!\!\left. +\sum_{m=1}^\infty c_m \frac{\sin(2 m z)}{z^{2m^2K+1}} \left(\sum_{n=0}^\infty \frac{c'_n}{z^{2 n}}\right)\right], 
\end{eqnarray} 
 where all the coefficients $a'_n$, $b_m$, $b'_n$, $c_m$ and $c'_n$ are 
 nonuniversal and need to be 
calculated by a fully microscopic theory (for possible methods see 
\textit{e.g.}~\cite{Amico_Korepin,Caux2007}). 
We note that Eq.~(\ref{eq:rhoGLL}): (i) generalizes 
Eq.~(\ref{eq:rhoLL}) and (ii) has the full structure of the exact 
result~(\ref{eq:rhoTG}) in the Tonks-Girardeau limit. As a direct 
 consequence of the series structure~(\ref{eq:rhoGLL}) the momentum 
 distribution will display singularities in its derivatives for $k=\pm 
 2 m \pi \rho_0$. 
 
\section{Regularized harmonic-fluid approach for bosons}

We proceed by outlining the method used. We describe the Bose gas with contact 
interactions (Lieb-Liniger model~\cite{LiebLiniger}) by a harmonic-fluid Hamiltonian, 
expressed in terms of the fields~$\theta(x)$ and~$\phi(x)$ which describe the density 
and phase fluctuations of the fluid~\cite{Haldane81}: 
\begin{equation} 
\label{eq:hamLL} 
\mathcal{H}_{LL}=\frac{\hbar v_s}{2\pi}\int\limits_0^L\mathrm{d}x\left[K\left(\nabla\phi(x)\right)^2+ 
\frac{1}{K}\left(\nabla\theta(x)\right)^2\right]. 
\end{equation} 
The parameters~$K$ and~$v_s$ entering Eq.~(\ref{eq:hamLL}) above are related to the 
microscopic interaction parameters~\cite{Cazalilla03}, the phase field~$\phi(x)$ is 
related to the velocity of the fluid $v(x)=\hbar \nabla \phi(x)/m$ and the 
field~$\theta(x)$ defines the fluctuations in the density profile. We have adopted here an 
effective low-energy description, which assumes that the collective excitations in the 
fluid are noninteracting and phonon-like. The description breaks down at a length 
scale~$a$ of the order of the typical inter-particle distance~$\rho_0^{-1}$, as it neglects the 
broadening and curvature of the Bose gas spectrum at finite 
momentum~\cite{LiebLiniger,Glazman}. Within the harmonic-fluid approximation, the bosonic field 
operator is expressed as $\Psi^\dag(x)=\sqrt{\rho(x)}\,\Exp{-i\phi(x)}$. Specifically, its 
expression in terms of the fields~$\theta(x)$ and~$\phi(x)$ 
reads~\cite{Haldane81,Cazalilla03} 
\begin{eqnarray} 
\label{eq:psi} 
\Psi^\dag(x)=\mathcal{A}\;[\rho_0+\Pi(x)]^{1/2} \sum_{m=-\infty}^{+\infty}\Exp{2mi\Theta(x)}\,\Exp{-i\phi(x)}, 
\end{eqnarray} 
where~$\mathcal{A}$ is a nonuniversal constant, $\Pi(x)=\nabla \theta(x)/\pi$, and for 
compactness of notation we have introduced the field $\Theta(x)=\theta_B+\pi\rho_0 
x+\theta(x)$, where~$\theta_B$ is chosen in order to ensure that the average of~$\theta(x)$ vanish. 
 
In order to calculate the one-body density matrix, we expand the fields~$\theta(x)$ 
and~$\phi(x)$ in terms of the normal modes (bosonic) operators~$b_j$, $b^\dagger_j$ which 
diagonalize the Hamiltonian~(\ref{eq:hamLL}) such that $\mathcal{H}_{LL}=\sum_j \hbar 
\omega_j b^\dagger_j b_j$, with $\omega_j=v_s k_j$.
As we are 
interested in the thermodynamic limit, we have chosen for simplicity the periodic 
boundary conditions for a Bose gas in a uniform box of length~$L$,
where $k_j=2\pi j/L$. 
In this case the mode 
expansion reads 
\begin{eqnarray} 
\phi(x)&=&\frac{1}{2\sqrt{K}}\sum_{j\neq0}\frac{\mathrm{sign}(k_j)\Exp{-a|k_j|/2}}{\sqrt{|j|}} 
\left(\Exp{i k_j x}b_j+\Exp{-i k_j x}b_j^\dag\right) 
\nonumber \\ &&+ \phi_0 + \frac{\pi x}{L}J, \label{eq:phi}\\ 
\Theta(x)&=&\frac{\sqrt{K}}{2}\sum_{j\neq0}\frac{\Exp{-a|k_j|/2}}{\sqrt{|j|}} 
\left(\Exp{i k_jx}b_j+\Exp{-ik_j x}b_j^\dag\right) 
\nonumber \\ && + \theta_0+\frac{\pi x}{L}N, \label{eq:theta} 
\end{eqnarray} 
with~$N$ and~$J$ being the particle number and angular momentum operators, and~$\phi_0$, 
$\theta_0$ being their conjugate fields in the phase-number representation~\cite{Haldane81}. 
The zero-mode fields~$\phi_0$, $\theta_0$ do not enter the calculation 
of the one-body density matrix, as it turns out to depend only on the differences 
$\theta(x_1)-\theta(x_2)$ and $\phi(x_1)-\phi(x_2)$. In the expressions~(\ref{eq:phi}) 
and~(\ref{eq:theta}) above we have introduced the short-distance cutoff $a\sim 
\rho_0^{-1}$, thus regularizing the effective theory. The one-body density matrix 
$\rho_1(x_1,x_2)=\langle \psi^\dagger(x_1)\psi(x_2)\rangle$ is obtained in the 
generalized harmonic-fluid approach from Eq.~(\ref{eq:psi}) as 
\begin{eqnarray}
\label{eq:rho1def}
&&\rho_1(x_1,x_2)\nonumber\\
&=&|\mathcal{A}|^2\!\!\!\sum_{(m,m')\in\mathbb{Z}^2}\!\!\!\langle[\rho_0+\Pi(x_1)]^{1/2} 
\,\Exp{i2m\Theta(x_1)}\Exp{-i\phi(x_1)}\nonumber\\
&&\phantom{|\mathcal{A}|^2\sum}\times
\Exp{i\phi(x_2)}\Exp{-i2m'\Theta(x_2)}[\rho_0+\Pi(x_2)]^{1/2} \rangle,
\end{eqnarray}
where the only nonvanishing leading terms satisfy~$m=m'$~\cite{Giamarchi_book}.

We detail now the calculation of the quantum average appearing in Eq.~(\ref{eq:rho1def}). 
In order to display its dependence only on differences between fields, we re-write the 
central term as 
\begin{eqnarray} 
\label{eq:central} 
&&\Exp{i2m\Theta(x_1)}\Exp{-i\phi(x_1)}\Exp{i\phi(x_2)}\Exp{-i2m\Theta(x_2)}\nonumber \\ 
&=&\Exp{i2m(\Theta(x_1)-\Theta(x_2))-i(\phi(x_1)-\phi(x_2))}\nonumber\\
&\times&\Exp{m\left[\theta(x_1)+\theta(x_2),\varphi(x_1)-\varphi(x_2)\right]}, 
\end{eqnarray} 
where $\varphi(x)=\phi(x)-\pi x\langle J \rangle/L$ and the commutator between 
the~$\theta$ and~$\phi$ fields is computed in Eq.~(\ref{eq:comm}) 
below~\footnote{In the following we will not consider the case of a macroscopic vorticity in the system, \textit{i.e.} we assume~$\langle J\rangle\ll N$.}.
We then perform a 
series expansion of the square root terms $[1+\Pi/\rho_0]^{1/2}$ in the one-body density 
matrix. We define $X=\Pi(x_1)/\rho_0$, $Y=\Pi(x_2)/\rho_0$, and 
$Z=i2m(\theta(x_1)-\theta(x_2))-i(\varphi(x_1)-\varphi(x_2))$. Using the fact that the 
fields~$X$, $Y$ and~$Z$ are Gaussian with zero average, we obtain from Wick's theorem 
\begin{eqnarray} 
\label{sqrtseries} 
&&\left\langle\sqrt{1+X}\,\Exp{Z}\sqrt{1+Y}\right\rangle\nonumber\\\nonumber
&=&\Exp{\frac{1}{2}\langle Z^2\rangle}\sum_{k,l=0}^\infty\sum_{j=0}^{\mathrm{min}(k,l)}
\frac{(2k)!(2l)!}{k!l!(2k-1)(2l-1)}\\\nonumber
&&\times\,
\frac{\langle X^2\rangle^\frac{k-j}{2}\langle Y^2\rangle^\frac{l-j}{2}\langle XY\rangle^j\!}{(i\sqrt{2})^{5(k+l)-2j}j!(k-j)!(l-j)!}\\
&&\times\,
\mathrm{H}_{k-j}\!\left(\frac{\langle XZ\rangle}{i\sqrt{2\langle X^2\rangle}}\right)
\mathrm{H}_{l-j}\!\left(\frac{\langle ZY\rangle}{i\sqrt{2\langle Y^2\rangle}}\right),
\end{eqnarray}
where $\mathrm{H}_n(x)$ are the Hermite polynomials. 
To second order in~$X$ and~$Y$ Eq.~(\ref{sqrtseries}) reads 
\begin{eqnarray} 
\left\langle\sqrt{1+X}\,\Exp{Z}\sqrt{1+Y}\right\rangle \simeq\Exp{\frac{1}{2}\langle 
 Z^2\rangle}\left[1+\frac{1}{2}\left(\langle XZ\rangle+\langle 
 ZY\rangle\right)\right. 
\nonumber \\ 
-\left.\frac{1}{8}\left(\langle X^2\rangle+\langle Y^2\rangle-2\langle XY\rangle\right)-\frac{1}{8}\left(\langle XZ\rangle-\langle ZY\rangle\right)^2\right]. 
\end{eqnarray} 
The main expression~(\ref{sqrtseries}) requires then the calculation of the various 
two-point correlation functions involving the three fields~$X$, $Y$, $Z$. All of them 
can be obtained from 
\begin{eqnarray} 
\!\!\!\!\langle\varphi(x_1)\varphi(x_2)\rangle&\!\!\!=\!\!\!&(\pi/L)^2\langle J_0^2\rangle x_1x_2-\frac{1}{4K}\ln C(x_1-x_2),\label{phiphi}\\ 
\!\!\!\!\langle\theta(x_1)\theta(x_2)\rangle&\!\!\!=\!\!\!&\pi\langle\Pi_0^2\rangle x_1x_2-\frac{K}{4}\ln 
C(x_1-x_2),\label{thetatheta} 
\\ 
\!\!\!\!\langle\theta(x_1)\varphi(x_2)\rangle&\!\!\!=\!\!\!&\frac{1}{4}\ln\!\left[\frac{1-\Exp{-2\pi 
  a/L-i2\pi(x_1-x_2)/L}}{1-\Exp{-2\pi 
  a/L+i2\pi(x_1-x_2)/L}}\right],\label{thetaphi} 
\\ 
\!\!\!\!\left[\theta(x_1),\varphi(x_2)\right]&\!\!\!=\!\!\!&i\frac{\pi}{L}(x_1-x_2)+2\langle\theta(x_1)\varphi(x_2)\rangle, 
\label{eq:comm} 
\end{eqnarray} 
where 
$C(x)=1-2\cos(2\pi x/L)\,\Exp{-2\frac{\pi a}{L}}+\Exp{-4\frac{\pi a}{L}}$, 
$\Pi_0=(N-\langle N\rangle)/L$, and $J_0=J-\langle J\rangle$. 
We are now in a position to calculate the correlators between the fields~$X$, $Y$, $Z$ in the thermodynamic limit 
($L\to\infty$, $N\to\infty$ at fixed $N/L=\rho_0$).
Using Eqs.~(\ref{phiphi})--(\ref{thetaphi}) we have 
\begin{eqnarray} 
\exp(\langle Z^2\rangle/2) 
&\simeq&\left(\frac{\alpha^2}{z^2+\alpha^2}\right)^{\frac{1}{4K}+Km^2} 
\\ 
\langle XZ\rangle=\langle ZY\rangle 
&\simeq&\frac{z}{2\alpha}\frac{z+i2Km\alpha}{z^2+\alpha^2} 
\\ 
\langle XY\rangle 
&\simeq&\frac{K}{2}\frac{\alpha^2-z^2}{(z^2+\alpha^2)^2} 
\\ 
\langle X^2\rangle=\langle Y^2\rangle&\simeq&\frac{K}{2\alpha^2}, 
\end{eqnarray} 
where $\alpha=\pi\rho_0 a$. 
Similarly, the commutator in Eq.~(\ref{eq:central}) is 
obtained from Eq.~(\ref{eq:comm}) as 
\begin{eqnarray} 
\exp\left(m\left[\theta(x_1)+\theta(x_2),\varphi(x_1)-\varphi(x_2)\right]\right) 
\nonumber \\ 
\simeq\left(\frac{a-i(x_1-x_2)}{a+i(x_1-x_2)}\right)^m 
\end{eqnarray} 
Notice that the effect of the zero-modes~$\Pi_0$ and~$J_0$ is absent in the 
thermodynamic limit, because it scales as~$1/L$. 
The series expansion in~$\Pi/\rho_0$ is valid for small fluctuations of the field~$\Pi(x)$ compared to the average density~$\rho_0$, $\sqrt{\langle X^2\rangle}\lesssim1$ \textit{i.e.} for $\alpha\gtrsim\sqrt{K/2}$.
By combining the previous equations we 
obtain the final result for the one-body density matrix in rescaled units, finding the structure displayed in Eq.~(\ref{eq:rhoGLL}): 
\begin{eqnarray} 
\rho_1(z)&=&\frac{\rho_\infty}{|z|^{1/2K}}\left[1+\frac{c_{0,2}}{z^2} +\frac{c_{0,4}}{z^4} 
+c_{1,2}\frac{\cos(2z)}{z^{2K}}\right.\nonumber\\ 
&&+\left. c_{1,4}\frac{\cos(2 z)}{z^{2K+2}}+c_{1,3}\frac{\sin(2z)}{z^{2K+1}}+...\right],
\label{rho1withPi} 
\end{eqnarray} 
with $\rho_\infty=|\mathcal{A}|^2 \alpha^{1/2K} c_{0,0}$.
 
\begin{figure} 
\centering 
\vspace{5mm} 
\includegraphics[width=0.35\textwidth]{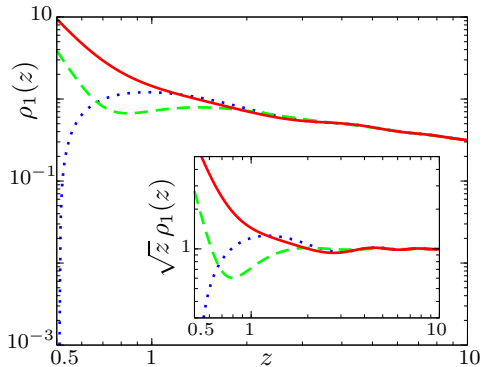} 
\caption{One-body density matrix in the Tonks-Girardeau limit~$K=1$ in units
 of~$\rho_0\rho_\infty$ as a function of the scaled relative
 coordinate $z=\pi\rho_0(x_1-x_2)$ (dimensionless). 
The result of the generalized harmonic-fluid approximation
 Eq.~(\ref{rho1withPi}) obtained without taking into
account the effect of the field~$\Pi(x)$  (solid line) is compared to the exact
 result Eq.~(\ref{eq:rhoTG}) (dashed line) and to the usual
 harmonic-fluid approximation Eq.~(\ref{eq:rhoLL}) (dotted
 line), with~$B_0=1$, $B_1=-1/2$ and $B_{m>1}=0$.
The value chosen for the cutoff parameter is $\alpha=1/2$. 
The inset shows the subleading behavior~$z^{1/2}\rho_1(z)$ of the
 one-body density matrix in the same notations and units as in the main graph.} 
\label{figrho1} 
\end{figure} 
 
A few comments are in order at this point. First, by taking the limit $a\to 0$ in 
Eqs.~(\ref{phiphi})--(\ref{eq:comm}) we recover the results of the standard harmonic-fluid 
approach, \textit{i.e.} Eq.~(\ref{eq:rhoLL}). Moreover, the generalized harmonic-fluid method produces also the corresponding coefficients of the series, namely to the order of approximation derived in this work, we have 
$c_{0,0} \simeq 1\,[ 1+\frac{1}{2\alpha}-\frac{K}{8\alpha^2}+\frac{K}{16\alpha^3} ]$; 
$c_{0,2} \simeq -\frac{\alpha^2}{4K} [ c_{0,0}+\frac{2K}{\alpha}-\frac{K^2}{2\alpha^2} ]/c_{0,0}$; 
$c_{0,4} \simeq \frac{(1+4K)\alpha^4}{32 K^2} [ 
 c_{0,0}+\frac{4K}{\alpha}+\frac{K^2}{\alpha^2}\frac{12K+1}{4K+1}-\frac{K^3}{\alpha^3}\frac{6}{4K+1} ]/c_{0,0}$;
$c_{1,2} \simeq -2 \alpha^{2K}$;
$c_{1,3} \simeq 4 \alpha^{2K+1} [
 c_{0,0}+\frac{K}{2\alpha}+\frac{K^2}{16\alpha^3} ]/c_{0,0}$;
$c_{1,4} \simeq \frac{(1+8K+4K^2)\alpha^{2K+2}}{ 2 K} [ 
 c_{0,0}+\frac{2K}{\alpha}\frac{1+4K}{1+8K+4K^2}+\frac{K^2}{2\alpha^2}\frac{1}{1+8K+4K^2}+\frac{K^3}{\alpha^3}\frac{1}{1+8K+4K^2} ]/c_{0,0}$.
However, it should be noted that these coefficients do not necessarily coincide \textit{e.g.} in the 
limit~$K=1$ with the exact ones in Eq.~(\ref{eq:rhoTG}). Our approach is still effective, 
as it suffers from some limitations: first of all, we have just used a single-parameter 
regularization which neglects the details of the spectrum of the Bose fluid. Second, our 
approach relies on a hydrodynamic-like expression for the field operator~(\ref{eq:psi}) 
which implicitly assumes that the fluctuations of the field~$\Pi(x)$ are ``small'', which 
is not always the case. 
On the other hand, we see from our derivation that the corrections due to 
the~$\Pi(x)$ fluctuations renormalize the coefficients of the series to all orders,
giving rise to the contributions in square brackets to the~$c_{ij}$ above,
but do not change the series structure.

Figure~\ref{figrho1} shows our results for the one-body density
matrix (Eq.~(\ref{rho1withPi}), solid lines) 
for~$K=1$ and a specific choice of the cutoff parameter $\alpha=1/2$ as
suggested by the analysis of other correlation 
functions~\cite{Didier09}. The comparison with the exact result for the
Tonks-Girardeau gas (Eq.~(\ref{eq:rhoTG}), dashed lines) yields a
reasonable agreement, the difference being due to the fact 
that we have used for the sake of illustration the explicit expression
for the coefficients $c_{i,j}$ derived in the current work. 
It should be noted in particular that 
 our generalized harmonic-fluid approach restores the correct trend at short distances as compared to the usual harmonic-fluid approach (Eq.~(\ref{eq:rhoLL}), dotted lines).

\section{Application to Fermions} 

It is possible to apply the above approach 
 to the case of a 1D spinless Fermi gas with odd-wave 
inter-particle interaction~\cite{CheShi99}, \textit{i.e.} the 1D analog of $p$-wave interactions. 
For any value of the coupling strength the (attractive)
Fermi gas can be mapped onto a (repulsive) Bose gas described by the
Lieb-Liniger model with dimensionless
coupling strength given by $\gamma_B=-1/\gamma_F$~\cite{CheShi99,Granger}, 
the spectrum of collective excitations being the same for the two models.
Hence, it is
possible to describe the interacting Fermi gas by the 
Luttinger-liquid Hamiltonian~(\ref{eq:hamLL}) as well,
except for the so-called Fermi-Tonks-Girardeau limit $\gamma_F=-\infty$ 
mapped onto the noninteracting Bose gas limit where the collective excitation 
spectrum is quadratic. 
 This allows 
 to estimate the fermionic one-body
 density matrix with a method similar to the one described above for
 the bosons, the
 fermionic nature of the particles being taken into account 
by a Jordan-Wigner transformation~\cite{Haldane81}
 on the bosonic field operator~(\ref{eq:psi}). 
In particular, we generalize the series expansion known from the usual harmonic-fluid approach, which in rescaled units reads (from~\cite{Haldane81}) 
\begin{equation}
\label{eq:rhoFLL}
\rho_{1F}^{\mathrm{LL}}(z)\sim \frac{1}{| 
 z|^{1/2K}}\sum_{m=0}^\infty C_m \frac{\sin(2 (m+1/2) z)}{z^{2(m+1/2)^2 K}}, 
\end{equation}
to the following series structure:
\begin{eqnarray}
\label{eq:rhoFGLL}
\rho_1(z)&\sim&
\frac{1}{|z|^{1/2K}}
\left[\sum_{m=0}^\infty d_m \frac{\sin((2m+1)z)}{z^{2(m+1/2)^2K}} 
\left(\sum_{n=0}^\infty\frac{d'_n}{z^{2 n}}\right)\right.\nonumber\\ 
&&+\left.\sum_{m=0}^\infty e_m\frac{\cos((2m+1)z)}{z^{2(m+1/2)^2K+1}}
\left(\sum_{n=0}^\infty \frac{e'_n}{z^{2 n}}\right)\right].
\end{eqnarray}
The coefficients in Eq.~(\ref{eq:rhoFGLL}) are nonuniversal and need a separate treatment.
The limit~$K=1$ is an exception though. 
It corresponds to the case of 
noninteracting fermions where the exact result for the one-body density
matrix $\rho_{1F}^{(0)}(x_1,x_2)=\sum_{|k|<k_F}\psi_k^*(x_1) \psi_k(x_2)$,
with~$\psi_k(x)$ being the single-particle orbitals, in the thermodynamic limit and in rescaled units reads 
\begin{equation}
\rho_{1F}^{(0)}(z)=\pi \rho_0 \frac{\sin(z)}{z}.
\end{equation} 
Hence, it turns out that for noninteracting fermions all the
coefficients of the series expression~(\ref{eq:rhoFGLL}) vanish except
the leading one.

\section{Conclusion}

In conclusion, we have developed a generalized harmonic-fluid
approximation and applied it to evaluate the large-distance behavior
 of the one-body density matrix both for one-dimensional interacting 
bosons and fermions 
 at arbitrary values of the interaction
strength -- the latter expressed in terms of the Luttinger 
parameter~$K$. 
In the case of
bosons in the Tonks-Girardeau limit~$K$=1 we recover 
 the full structure of the series expansion known for the exact
 result.
In the case of noninteracting fermions, the exact result shows
that the series has only one nonvanishing term.
In perspective, it could be interesting to estimate the coefficients
of the series expansion at arbitrary interaction strength.

\begin{acknowledgments} 
We acknowledge discussions with D.M.~Gangardt, L.I.~Glazman, R.~Santachiara and M.B.~Zvonarev. We thank IUF, CNRS and the MIDAS-STREP project for financial support. 
\end{acknowledgments}

\end{document}